\documentclass[11pt]{article}
\usepackage{amssymb,amsmath}
\usepackage{graphicx,rotate,subfigure}
\usepackage{cite,url}

\usepackage{color}

\def \order(#1){{\cal O} \left(#1 \right)}

\evensidemargin=\oddsidemargin
\let\CAptIOn=\caption
\def\caption#1{\CAptIOn{\small\em #1}}

\def\Eqn#1{Eq.\ (\ref{#1})}
\def\Eqs#1#2{Eqs.\ (\ref{#1}) and (\ref{#2})}

\begin{document}




\begin{center}
{\Large \bf Scalar sector properties of two-Higgs-doublet models\\ with a
  global U(1) symmetry } \\ 
\vspace*{1cm}  {\sf
    Gautam Bhattacharyya$^1$, Dipankar Das$^1$, Palash B. Pal$^1$,
    M. N. Rebelo$^2$} 
\\
\vspace{10pt} $^1${\em Saha Institute of Nuclear
    Physics, 1/AF Bidhan Nagar, Kolkata 700064, India}

\vspace{10pt} $ ^2${\em Centro de F\'{i}sica Te\'{o}rica de
  Part\'{i}culas, Instituto Superior T\'ecnico, \\ Universidade de
  Lisboa, Av. Rovisco Pais, 1049-001 Lisboa, Portugal} 
\normalsize
\end{center}

\begin{abstract}

We analyze the scalar sector properties of a general class of
two-Higgs-doublet models which has a global U(1) symmetry in the
quartic terms.  We find constraints on the parameters of the potential
from the considerations of unitarity of scattering amplitudes, the
global stability of the potential and the $\rho$-parameter.  We
concentrate on the spectrum of the non-standard scalar masses in the
decoupling limit which is preferred by the Higgs data at the LHC.  We
exhibit charged-Higgs induced contributions to the diphoton decay
width of the 125\,GeV Higgs boson and its correlation with the
corresponding $Z\gamma$ width.

\end{abstract}

\bigskip

\section{Introduction}
A new boson with a mass of nearly 125\,GeV has been observed by the
CMS and ATLAS collaborations at the LHC \cite{Aad:2012tfa,
  Chatrchyan:2012ufa}.  It is too early to comment on its spin.
Preliminary analyses hint towards a $0^+$ state
\cite{Chatrchyan:2012jja}.  Different production mechanisms and decay
rates of this new scalar are compatible with what we expect for the
Standard Model (SM) Higgs boson.  Currently it is a subject of intense
study \cite{Ferreira:2011aa, Ferreira:2012my, Swiezewska:2012eh,
  Drozd:2012vf, Chang:2012ve, Craig:2013hca, Chiang:2013ixa,
  Jiang:2013pna, Chen:2013rba, Eberhardt:2013uba, Basso:2013wna,
  Swiezewska:2013uya} whether the observed scalar is alone or is a
part of a richer scalar sector.  The simplest scenario entertaining
this possibility is the class of two-Higgs-doublet models (2HDMs)
\cite{Branco:2011iw}.  In these extensions, the value of the
electroweak $\rho$-parameter remains unity at tree level as in the SM.

A general problem that one encounters with 2HDMs is the presence of
Higgs-mediated flavor changing neutral currents (FCNCs).  To ensure
natural flavor conservation, one usually relies on a symmetry that
completely eliminates tree level Higgs mediated FCNC.  One example is
the imposition of a discrete $Z_2$ symmetry, assigned in such a way
that all fermions with the same electric charge obtain their masses
from the vacuum expectation value (VEV) of one particular scalar
doublet \cite{Glashow:1976nt, Paschos:1976ay, Grossman:1994jb}.  A
larger symmetry of this kind is a global U(1) symmetry, employed in a
different context in Peccei-Quinn model.  Alternatively, one can admit
tree level Higgs mediated FCNCs which are suppressed by small factors
involving small entries of the CKM matrix \cite{Joshipura:1990pi,
  Antaramian:1992ya, Hall:1993ca}.  Such a model with a continuous
global U(1) symmetry was proposed Branco, Grimus and Lavoura
\cite{Branco:1996bq}, which and its later generalizations
\cite{Botella:2009pq, Botella:2011ne} fall under the category of
models with minimal flavor violation \cite{D'Ambrosio:2002ex}.

We however concentrate only on the scalar sector with a global U(1)
symmetry, regardless of the transformation properties of the fermions
under this symmetry.  Thus our observations extend well beyond the
scope of any such individual flavor model.  Ferreira and Jones
\cite{Ferreira:2009jb} have done a thorough analysis of stability and
perturbativity bounds on the parameter space of such models together
with the consideration of non-observation of any scalar at LEP.  Our
motivation in this paper is to explore the scalar potential with a
softly broken U(1) symmetry in view of the observation of a 125\,GeV
Higgs boson ($h$) at the LHC, using constraints from unitarity of
scalar scattering cross-sections, stability of the potential and
electroweak precision tests.  These considerations restrict the
spectrum of the non-standard scalars.  Since the LHC Higgs data seem
to be compatible with the SM expectations, we restrict ourselves to
the decoupling limit, explained later.  In this limit, we study the
charged-Higgs induced contribution to $h\to\gamma\gamma$ decay width
and its correlation to $h \to Z\gamma$.

\section{Bounds on masses from stability and
  unitarity} \label{s:potential}
We take the scalar potential as follows\cite{HHGuide}:
\begin{eqnarray}
 V(\phi_1, \phi_2) &=& 
 \lambda_1 \left( \phi_1^\dagger\phi_1 - \frac{v_1^2}{2} \right)^2 
+\lambda_2 \left( \phi_2^\dagger\phi_2 - \frac{v_2^2}{2} \right)^2 
\nonumber \\*
&& +\lambda_3 \left( \phi_1^\dagger\phi_1 + \phi_2^{\dagger}\phi_2 
- \frac{v_1^2+v_2^2}{2} \right)^2
+\lambda_4 \left(
(\phi_1^{\dagger}\phi_1) (\phi_2^{\dagger}\phi_2) -
(\phi_1^{\dagger}\phi_2) (\phi_2^{\dagger}\phi_1)
\right) 
\nonumber \\*
&& + \lambda_5 \left( \frac12 \Big( \phi_1^\dagger\phi_2 +
\phi_2^\dagger\phi_1 - v_1v_2 \Big) \right)^2 
+ \lambda_6 \left( \frac1{2i} \Big( \phi_1^\dagger\phi_2 -
\phi_2^\dagger\phi_1 \Big) \right)^2 \,,
\label{potential}
\end{eqnarray}
where the $\lambda$'s are real because of hermiticity of the
Lagrangian.  All terms in the potential are invariant under the
discrete symmetry:
\begin{eqnarray}
\phi_1 \to \phi_1 \,, \qquad
\phi_2 \to - \phi_2 \,,
\label{dsymm}
\end{eqnarray}
except a term 
\begin{eqnarray}
- \frac12 \lambda_5 v_1 v_2 (\phi_1^\dagger\phi_2 + \phi_2^\dagger\phi_1)
\label{soft}
\end{eqnarray}
which breaks the symmetry only softly.  We use the following
parametrization for the Higgs doublets:
\begin{eqnarray}
  \phi_i(x) = \begin{pmatrix}
             w_i^+(x) \\
             \frac{1}{\sqrt2}(v_i+h_i(x)+iz_i(x)) 
               \end{pmatrix} \,,
\label{phi}
\end{eqnarray}
where the $v_i$ denote the vacuum expectation values (VEVs) of the two
doublets, assumed to be real and positive without any loss of
generality.  It is useful to define the following parameters,
\begin{subequations}
 \begin{eqnarray}
  \tan \beta &=& \frac{v_2}{v_1}  \qquad (0 \leq \beta \leq
  \frac{\pi}{2}) \,, \\* 
 v &=& \sqrt{v_1^2 + v_2^2} = 246 ~ {\rm GeV} \,.
\label{v}
\end{eqnarray}
\end{subequations}
There will be five physical Higgs bosons in this model.  We denote the
charged ones by $\xi^\pm$, the CP-odd one by $A$, and use the symbols
$H$ and $h$ to denote the heavy and light CP even Higgs bosons
respectively.  There will also be the combinations $\omega^\pm$, $\zeta$
which are the three would-be Goldstone bosons eaten by the gauge
bosons.  These combinations are given by
\begin{eqnarray}
 \begin{pmatrix}
             \omega^\pm \\ \xi^\pm 
 \end{pmatrix} &=&
  \begin{pmatrix}
             c_\beta & s_\beta \\
             -s_\beta & c_\beta 
 \end{pmatrix}
  \begin{pmatrix}
             w_1^\pm \\ w_2^\pm 
 \end{pmatrix}, \\ 
 \begin{pmatrix}
             \zeta \\ A 
 \end{pmatrix} &=&
  \begin{pmatrix}
             c_\beta & s_\beta \\
             -s_\beta & c_\beta 
 \end{pmatrix}
  \begin{pmatrix}
             z_1 \\ z_2
 \end{pmatrix}, 
\label{zetaA}\\ 
 \begin{pmatrix}
             H \\ h 
 \end{pmatrix} &=&
  \begin{pmatrix}
             c_\alpha & s_\alpha \\
             -s_\alpha & c_\alpha 
 \end{pmatrix}
  \begin{pmatrix}
             h_1 \\ h_2 
 \end{pmatrix} \,,
\label{defalpha}
\end{eqnarray}
where $c_\alpha \equiv \cos\alpha$, $s_\alpha\equiv \sin\alpha$,
and likewise for $\beta$.  The angle of rotation for the matrix in
\Eqn{defalpha} is given by
\begin{eqnarray}
 \tan 2\alpha =
 \frac{2 (\lambda_3 +\frac14 \lambda_5) v_1v_2} {\lambda_1v_1^2 -
   \lambda_2v_2^2 + (\lambda_3 - \frac14 \lambda_5)(v_1^2-v_2^2)}
 \,,  \qquad (-\frac{\pi}{2} \leq 2\alpha \leq \frac{\pi}{2}) \,.
\end{eqnarray}

We have $v_1$, $v_2$, and six $\lambda$-parameters as independent
parameters in the potential of \Eqn{potential}.  They are related to
the four physical Higgs boson masses by the following relations:
\begin{subequations}
 \label{lambdas}
\begin{eqnarray}
 \lambda_1 &=& \frac{1}{2v^2c_\beta^2}\left[c_\alpha^2m_H^2
   +s_\alpha^2m_h^2-\frac{s_\alpha
     c_\alpha}{\tan\beta}(m_H^2-m_h^2)\right]
 -\frac{\lambda_5}{4}(\tan^2\beta-1)  \,, \\
\lambda_2 &=& \frac{1}{2v^2s_\beta^2}\left[s_\alpha^2m_H^2
  +c_\alpha^2m_h^2 - s_\alpha c_\alpha\tan\beta(m_H^2-m_h^2)\right]
-\frac{\lambda_5}{4}\left(\frac{1}{\tan^2\beta}-1\right)  \,, \\
\lambda_3 &=& \frac{1}{2v^2}\frac{s_\alpha c_\alpha}{s_\beta
  c_\beta}(m_H^2-m_h^2) -\frac{\lambda_5}{4} \,, \\
\lambda_4 &=& \frac{2}{v^2}m_\xi^2 \,, \\
\lambda_6 &=& \frac{2}{v^2}m_A^2 \,.
\label{lam5}
\end{eqnarray}
\end{subequations}
Thus an alternative way of counting the independent parameters is
through the four masses, the two angles $\alpha$ and $\beta$, the
electroweak VEV $v$, and the parameter $\lambda_5$, which appear on
the right-hand sides of \Eqn{lambdas}.  In this set of eight
parameters, $v$ is known as in \Eqn{v}, and so is the lightest CP-even
Higgs mass as 125\,GeV.

The other thing that we will need from the scalar potential is the
cubic coupling involving $h\xi^+\xi^-$, denoted by $g_{h\xi\xi}$,
which will be employed in the $h \to \gamma\gamma$ and $h \to Z
\gamma$ decay rates of the Higgs boson.  This coupling is given by
\begin{eqnarray}
g_{h\xi\xi} = v \Big[ (\lambda_1 + \lambda_2) s_\alpha s_\beta
  s_{2\beta} - \lambda_2 s_{2\beta} c_{\beta-\alpha} - (2\lambda_3 +
  \lambda_4) s_{\beta-\alpha} + \frac12 \lambda_5 s_{2\beta}
  c_{\alpha+\beta} \Big] \,.
\label{ghxixi}
\end{eqnarray}
In terms of physical masses, this coupling can be written
as~\cite{Djouadi:1996yq} 
\begin{eqnarray}
g_{h\xi\xi} = - \frac 1v \Big[ (2m_h^2 - \lambda_5 v^2) \;
  {\cos(\alpha+\beta) \over \sin 2\beta}
  + (2 m_\xi^2 - m_h^2) \sin (\beta-\alpha) \Big] \,.
\end{eqnarray}

In the present work, we explore the consequences of a global symmetry
that is larger than that given in \Eqn{dsymm}:
\begin{eqnarray}
\phi_1 \to \phi_1 \,, \qquad
\phi_2 \to e^{i\theta} \phi_2 \,.
\label{U1}
\end{eqnarray}
On the quartic terms of the scalar potential, this symmetry is
realized by putting 
\begin{eqnarray}
\lambda_5  = \lambda_6 \,,
\label{BGL}
\end{eqnarray}
which means that the potential now reads
\begin{eqnarray}
 V(\phi_1, \phi_2) &=& 
 \lambda_1 \left( \phi_1^\dagger\phi_1 - \frac{v_1^2}{2} \right)^2 
+\lambda_2 \left( \phi_2^\dagger\phi_2 - \frac{v_2^2}{2} \right)^2 
+\lambda_3 \left( \phi_1^\dagger\phi_1 + \phi_2^{\dagger}\phi_2 
- \frac{v_1^2+v_2^2}{2} \right)^2
\nonumber \\ 
 &&+\lambda_4 \left(
(\phi_1^{\dagger}\phi_1) (\phi_2^{\dagger}\phi_2) -
(\phi_1^{\dagger}\phi_2) (\phi_2^{\dagger}\phi_1)
\right) 
+\lambda_5 \left| \phi_1^\dagger\phi_2 - \frac{v_1v_2}{2} \right|^2 \,. 
\label{BGLpot}
\end{eqnarray}
The term of \Eqn{soft}, which breaks the discrete symmetry softly, is
a soft explicit breaking term of the global U(1) symmetry as well.  In
the limit $\lambda_5=0$, the U(1) symmetry is exact in the potential,
and the spontaneous breaking of it through $v_2 \neq0$, since only
$\phi_2$ undergoes a nontrivial U(1) phase transformation, gives rise
to the Goldstone boson $A$. Since $\lambda_5 \to 0$ corresponds to an
enhanced symmetry, it is expected to remain small {\em a la} 't Hooft,
and consequently, the CP-odd $A$ can remain naturally light.  We will
later see under what conditions we can entertain a light $A$ boson,
and what are its consequences.

Conditions for the potential being bounded from below were examined
for more general potentials in 2HDM \cite{Sher:1988mj,
  Gunion:2002zf}.  Using the relation of \Eqn{BGL}, these conditions
read
\begin{subequations}
\label{vacuum}
\begin{eqnarray}
&& \lambda_1 + \lambda_3 > 0 \,, 
\label{vac12}\\*
&& \lambda_2 + \lambda_3  > 0 \,, 
\label{vac23} \\
&& 2\lambda_3 + \lambda_4 + 
2\sqrt{ (\lambda_1+\lambda_3) (\lambda_2+\lambda_3) } > 0 \,, \\*
 && 2\lambda_3 + \lambda_5 + 
2\sqrt{ (\lambda_1+\lambda_3) (\lambda_2+\lambda_3) } > 0 \,.
\end{eqnarray}
\end{subequations}
While the conditions in \Eqn{vacuum} put lower bounds on certain
combinations of the quartic couplings, there exist upper bounds on
these couplings arising from the consideration of perturbative
unitarity \cite{Lee:1977eg}.  Scattering amplitudes involving
longitudinal gauge bosons and Higgs bosons comprise the elements of an
$S$-matrix, having 2-particle states as rows and columns.  The
eigenvalues of this matrix are restricted by $|a_0|<1$, where $a_0$ is
the $l=0$ partial wave amplitude.  These conditions translate into
upper limits on combinations of Higgs quartic couplings, which for
multi-Higgs models have been derived by different authors
\cite{Maalampi:1991fb, Kanemura:1993hm, Akeroyd:2000wc,
  Horejsi:2005da}.  Imposing the condition of \Eqn{BGL}, these
constraints assume the following form:
\begin{subequations}
 \label{unitarity}
\begin{eqnarray}
 && \Big| 2\lambda_3 - \lambda_4 + 2\lambda_5 \Big| \leq 16\pi \,,  \\*
 && \Big| 2\lambda_3 + \lambda_4 \Big| \leq 16\pi \,, 
\label{uni2}
\\
 && \Big| 2\lambda_3 + \lambda_5 \Big| \leq 16\pi \,, \\
 && \Big| 2\lambda_3 + 2\lambda_4 - \lambda_5 \Big| \leq 16\pi \,, \\
 && \Big| 3(\lambda_1 + \lambda_2 + 2\lambda_3) \pm 
\sqrt{9(\lambda_1-\lambda_2)^2 + (4\lambda_3 + \lambda_4 +
  \lambda_5)^2}\Big| \leq 16\pi \,, 
\label{uni5}
\\ 
 && \Big|(\lambda_1+\lambda_2+2\lambda_3) \pm
\sqrt{(\lambda_1-\lambda_2)^2 + (\lambda_4 - \lambda_5)^2}\Big| \leq
  16\pi \,, \\ 
 && \Big|(\lambda_1+\lambda_2+2\lambda_3) \pm
  (\lambda_1-\lambda_2) \Big| \leq 16\pi \,.
\label{uni7}
\end{eqnarray}
\end{subequations}

\begin{figure}[t]
\rotatebox{90}{\quad\quad\quad\quad$m_A$ (GeV)}
\includegraphics[scale=0.64]{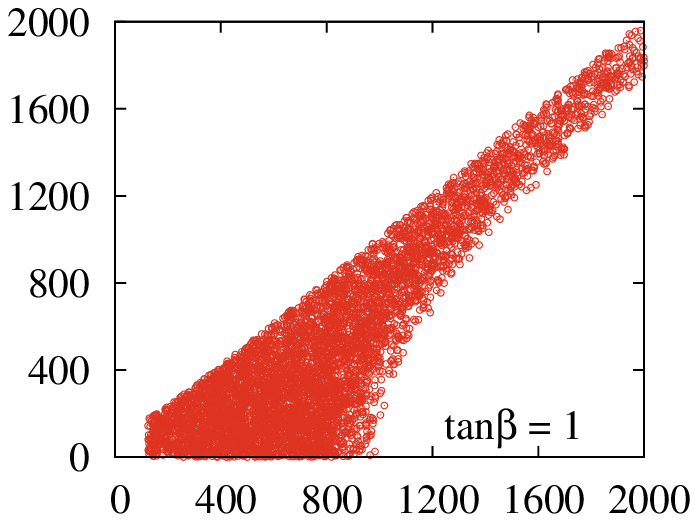} 
\includegraphics[scale=0.64]{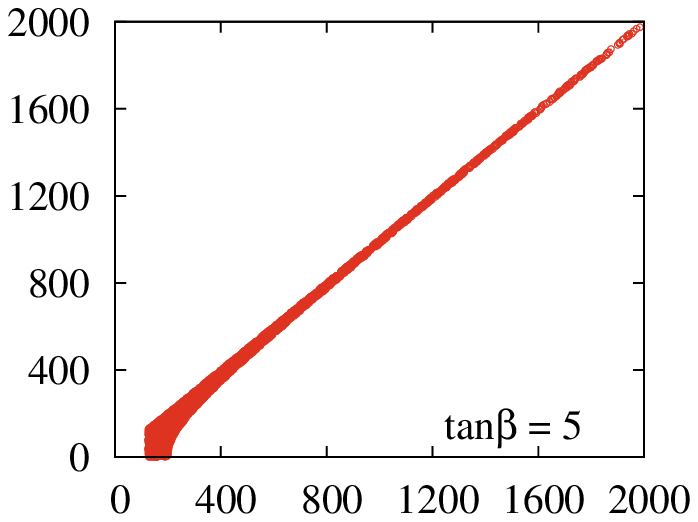}
\includegraphics[scale=0.64]{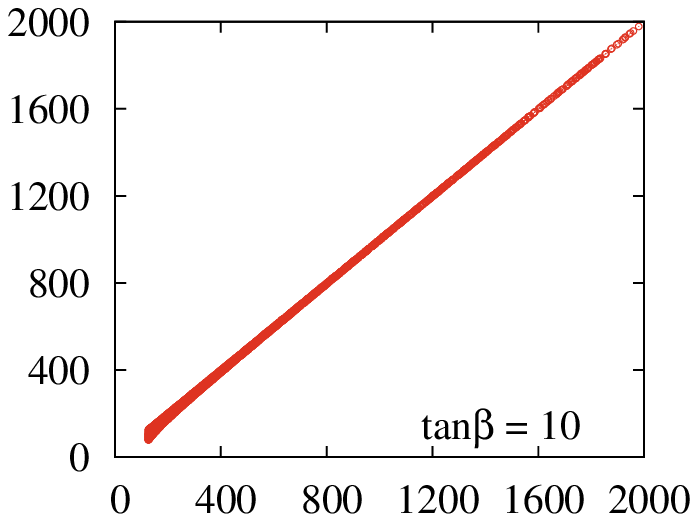}
\centerline{ \null \hfill $m_H$ (GeV) \quad}

\rotatebox{90}{\quad\quad\quad\quad$m_\xi$ (GeV)}
\includegraphics[scale=0.64]{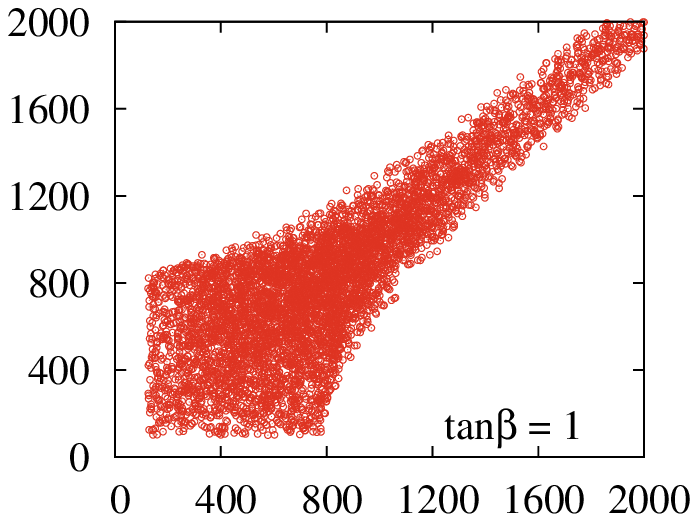} 
\includegraphics[scale=0.64]{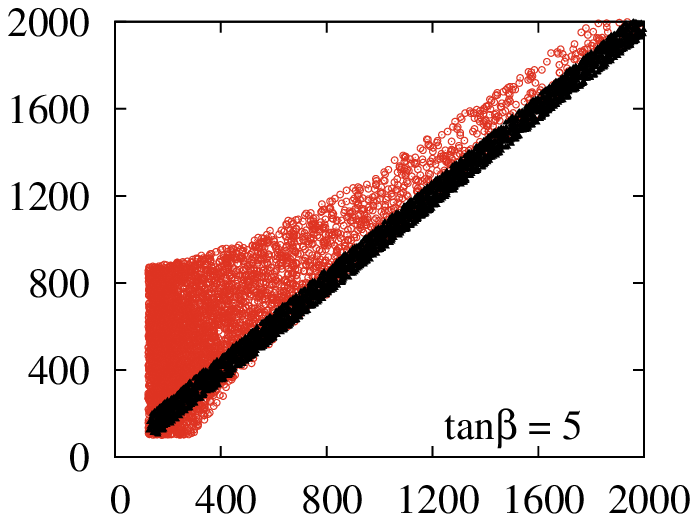}
\includegraphics[scale=0.64]{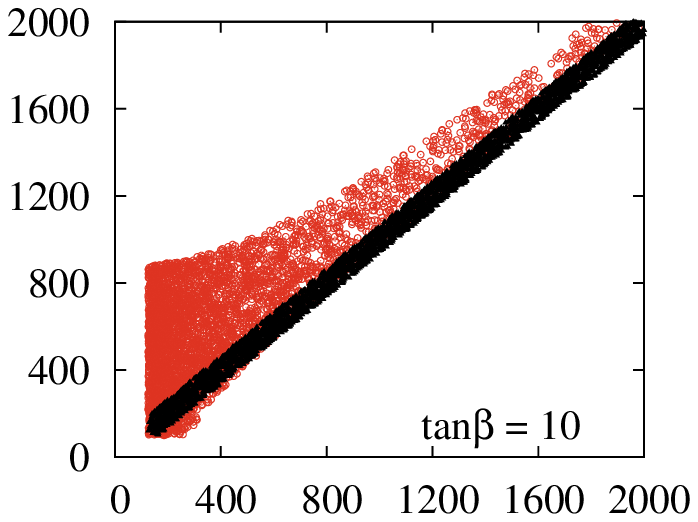}
\centerline{ \null \hfill $m_H$ (GeV) \quad}

\rotatebox{90}{\quad\quad\quad\quad$m_\xi$ (GeV)}
\includegraphics[scale=0.64]{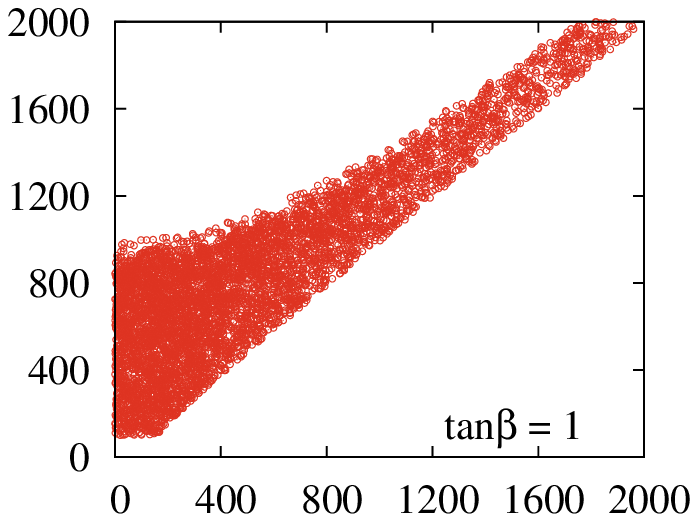} 
\includegraphics[scale=0.64]{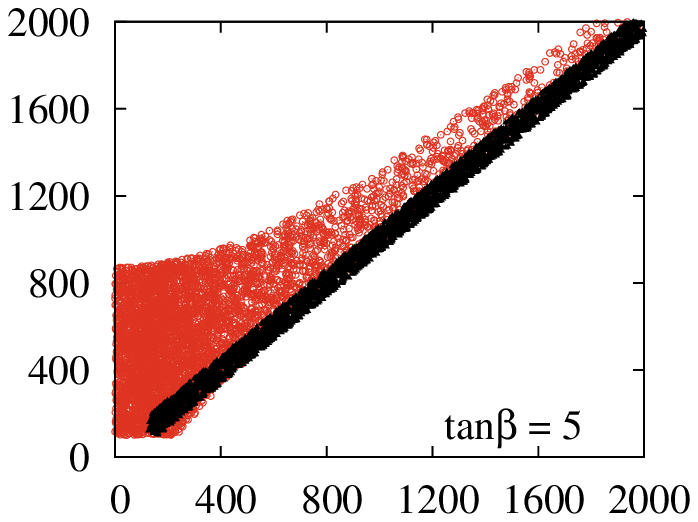}
\includegraphics[scale=0.64]{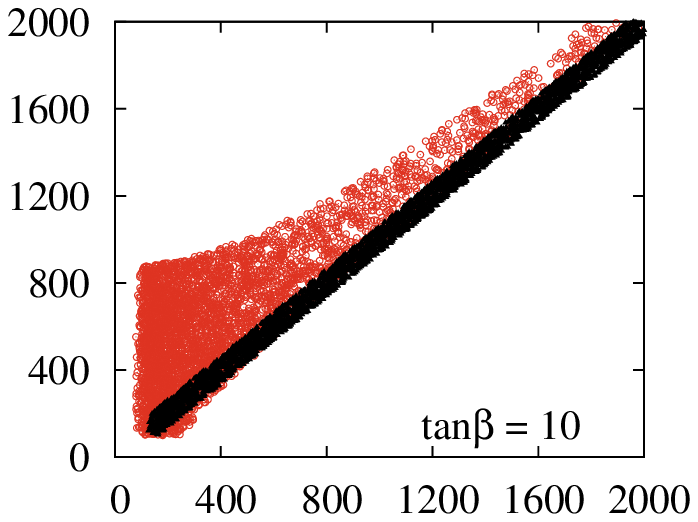}
\centerline{ \null \hfill $m_A$ (GeV) \quad}

\caption{Regions allowed in $m_H$-$m_A$, $m_H$-$m_\xi$ and
  $m_A$-$m_\xi$ planes from unitarity and stability (red points), and
  from $T$-parameter (black points), for three choices of
  $\tan\beta$.  We put $m_\xi > 100$\,GeV and $m_H > m_h$.}
\label{f:planes}
\end{figure}

It is worth noting \cite{Branco:1996bq} at this stage that if we
rotate the basis $h_1$-$h_2$ by the same angle $\beta$ which appears
in \Eqn{zetaA}, we obtain the states
\begin{eqnarray}
\begin{pmatrix}
             H^0 \\  R
 \end{pmatrix} =
  \begin{pmatrix}
             c_\beta & s_\beta \\
             -s_\beta & c_\beta 
 \end{pmatrix}
  \begin{pmatrix}
             h_1 \\  h_2
 \end{pmatrix} \,,
\end{eqnarray}
which has the property that $H^0$ has the exact SM couplings with the
fermions and gauge bosons.  The state $R$ does not have any cubic
gauge coupling at the tree level.  It can however have flavor changing
Yukawa couplings.  The lighter CP even mass eigenstate, $h$, is
related to $H^0$ and $R$ via the transformation
\begin{eqnarray}
 h = \sin(\beta-\alpha) H^0 + \cos(\beta-\alpha) R \,.
\end{eqnarray}
If it is eventually settled that the Higgs boson observed at the LHC
has SM-like gauge and Yukawa couplings, then we will require
\begin{eqnarray}
\sin(\beta-\alpha) \approx 1 \,,
\end{eqnarray}
which has been referred to as the {\em decoupling limit}
\cite{Gunion:2002zf}.  Unless otherwise stated, we make  the
following assumptions in all our subsequent analysis:
\begin{itemize}
\item $\beta-\alpha=\pi/2$;
\item \Eqn{BGL} holds;
\item $m_h = 125\,$GeV;
\item $m_\xi > 100$\,GeV, which is a rough lower bound from direct
  searches \cite{PDG}. 
\end{itemize}
The resulting constraints on the scalar masses imposed by the
stability and unitarity conditions in \Eqs{vacuum}{unitarity} have
been plotted in Fig.~\ref{f:planes} for $\tan\beta=1$, 5 and 10 by
performing random scan over all non-standard scalar masses.

We note at this point that the splitting between the heavy scalar
masses is also constrained by the oblique electroweak $T$-parameter.
In the present case, the expression of the $T$-parameter in the
decoupling limit is given by~\cite{He:2001tp, Grimus:2007if}
\begin{eqnarray}
T = {1 \over 16\pi \sin^2 \theta_w M_W^2} \Big[ F(m_\xi^2,m_H^2) +
  F(m_\xi^2, m_A^2) - F(m_H^2,m_A^2) \Big] \,,
\label{T}
\end{eqnarray}
with
\begin{eqnarray}
F(x,y) = {x+y \over 2} - {xy \over x-y} \; \ln (x/y) \,.
\end{eqnarray}
Taking the new physics contribution to the $T$-parameter
as~\cite{Baak:2013ppa} 
\begin{eqnarray}
T = 0.05 \pm 0.12 \,,
\end{eqnarray}
we project the $2\sigma$ constraints in Fig.~\ref{f:planes} for
$\tan\beta=5$ and 10.  The following salient features emerge from the
plots.
\begin{enumerate}
\item\label{i:degen} There is a correlation between $m_A$ and $m_H$
  which gets stronger for larger values of $\tan\beta$, to the extent
  that they become nearly degenerate once $\tan\beta$ crosses 10.  To
  understand this, we note that \Eqs{vac12}{vac23}, along with
  \Eqn{uni5} for example, imply the inequality
\begin{eqnarray}
0 \leq \lambda_1 + \lambda_2 + 2\lambda_3 \leq {16\pi \over 3} \,.
\end{eqnarray}
In terms of the scalar masses in \Eqn{lambdas}, this reads
\begin{eqnarray}
0 \leq (m_H^2-m_A^2)(\tan^2\beta+\cot^2\beta)+2m_h^2 
\leq {32\pi v^2 \over 3} \,.
\label{uniM1}
\end{eqnarray}
Clearly, for $\tan\beta$ away from unity, $H$ and $A$ are almost
degenerate.

\item There is a similar correlation between $m_H$ and $m_\xi$, but
  this time without any dependence on $\tan\beta$.  This can again be
  seen from the inequality
\begin{eqnarray}
\Big| 2m_\xi^2 - m_H^2 - m_A^2 + m_h^2 \Big| \leq 16\pi v^2 \,,
\label{uniM2}
\end{eqnarray}
which follows from \Eqn{uni2}.

\item As regards the non-standard scalars, the unitarity conditions
  essentially apply on the difference of their squared masses.  Thus,
  any individual mass can be arbitrarily large without affecting the
  unitarity conditions.  This conclusion crucially depends on the
  existence of a U(1) symmetry of the potential.  When the symmetry of
  the potential is only a discrete $Z_2$, considerations of unitarity
  do restrict the individual non-standard
  masses~\cite{Kanemura:1993hm, Akeroyd:2000wc, Horejsi:2005da}.

\item To provide an intuitive feel on the constraints from the
  $T$-parameter, we assume $m_H=m_A$, which is anyway dictated by the
  unitarity constraints for $\tan\beta$ somewhat away from unity.  It
  then follows from \Eqn{T} that the splitting between $m_\xi$ and
  $m_H$ is approximately 50\,GeV, for $|m_\xi-m_H| \ll m_\xi,m_H$.  It
  turns out from Fig.~\ref{f:planes} that the constraints from the
  $T$-parameter are stronger than that from unitarity and stability.

  For $\tan\beta=1$, unitarity and stability do not compel $m_H$ and
  $m_A$ to be very close.  In this case, the $T$-parameter cannot give
  any definitive constraints in the planes of the heavy scalar masses,
  unlike the unitarity and stability constraints.  For this reason, we
  have shown only the latter constraints in Fig.~\ref{f:planes} for
  $\tan\beta=1$.

\item \Eqn{T} shows that the contribution to the $T$-parameter from
  scalar loops is vanishingly small irrespective of the value of $m_A$
  as long as $m_H \approx m_\xi$.  As Fig.~\ref{f:planes} shows, even
  $m_A=0$ is allowed for $\tan\beta=1$ and $\tan\beta=5$.  A light
  pseudoscalar is experimentally allowed, and various aspects of its
  phenomenology in the context of 2HDM have been discussed in the
  literature~\cite{Larios:2001ma, Larios:2002ha, Cervero:2012cx}.  The
  value $\tan\beta=10$ is already large enough so that the conclusion
  of item \ref{i:degen} applies, and $m_H$ drags $m_A$ with it.

\item Thus, for moderate or large $\tan\beta$, the unitarity and
  stability constraints, together with the constraints coming from the
  $T$-parameter, imply that all three heavy scalar states are nearly
  degenerate in the decoupling limit.

\end{enumerate}

\section{Modifications in Higgs decay width}\label{s:decay}
Since we are working in the decoupling limit, the couplings of $h$
with the fermions and gauge bosons will be exactly like in the SM.
The production cross section of $h$ will therefore be as expected in
the SM.  All the tree level decay widths of $h$ will also have the SM
values for the same reason.  Loop induced decays like $h\to
\gamma\gamma$ and $h\to Z\gamma$ will however have additional
contributions from virtual charged scalars ($\xi^\pm$).  Since the
branching fractions of such decays are tiny, the total decay width is
hardly modified.

The contribution of the $W$-boson loop and the top loop diagrams to
$h\to \gamma\gamma$ and $h\to Z\gamma$ are same as in the SM.  As
regards the charged scalar induced loop, we first parametrize the
cubic coupling $g_{h\xi\xi}$, given in \Eqn{ghxixi}, in the following
way: 
\begin{eqnarray}
g_{h\xi\xi} =  \kappa \; \frac{g m_{\xi}^2}{M_W} \,,
\label{defkappa}
\end{eqnarray}
where $\kappa$ is dimensionless.  The diphoton decay width is then
given by \cite{Djouadi:2005gi, Djouadi:2005gj}:
\begin{eqnarray}
 \Gamma (h\to \gamma\gamma) = \frac{\alpha^2g^2}{2^{10}\pi^3}
 \frac{m_h^3}{M_W^2} \Big|F_W + \frac{4}{3}F_t  + \kappa F_\xi \Big|^2
 \,, 
\label{h2gg}
\end{eqnarray}
where, introducing the notation
\begin{eqnarray}
\tau_x \equiv (2m_x/m_h)^2 \,,
\end{eqnarray}
the values of $F_W$, $F_t$ and $F_{\xi}$ are given by
\begin{subequations}
\begin{eqnarray}
 F_W &=& 2+3\tau_W+3\tau_W(2-\tau_W)f(\tau_W) \,,  \\
 F_t &=& -2\tau_t \big[1+(1-\tau_t)f(\tau_t)\big] \,,  \\
 F_\xi &=& -\tau_\xi \big[ 1-\tau_{\xi}f(\tau_{\xi}) \big] \,.
\end{eqnarray}
\end{subequations}
With our assumptions about the masses declared earlier, $\tau_x >1$
for $x=W$, $t$, $\xi^\pm$.   Then 
\begin{eqnarray}
f(\tau) =
\left[\sin^{-1}\left(\sqrt{1/\tau}\right)\right]^2 \,.
\label{f}
\end{eqnarray}
The decay width for $h\to Z\gamma$ can analogously be written as:
\begin{eqnarray}
 \Gamma (h\to Z\gamma) = \frac{\alpha^2g^2}{2^{9}\pi^3}
 \frac{m_h^3}{M_W^2} \Big|A_W + A_t  + \kappa A_{\xi}\Big|^2
 \left(1-\frac{M_Z^2}{m_h^2}\right)^3 \,,
\label{h2Zg}
\end{eqnarray}
where, introducing
\begin{eqnarray}
\eta_x = (2m_x/M_Z)^2 \,,
\end{eqnarray}
the values of $A_W$, $A_t$ and $A_{\xi}$ are given by~\cite{HHGuide}
\begin{subequations}
\begin{eqnarray}
 A_W &=& \cot \theta_w\bigg[ 4(\tan^2\theta_w - 3)I_2(\tau_W,\eta_W)
 \nonumber \\*
&& \null +\bigg\{ \left(5+\frac{2}{\tau_W}\right) -
 \left(1+\frac{2}{\tau_W}\right)\tan^2\theta_w \bigg\}
 I_1(\tau_W,\eta_W)\bigg] \,, 
 \\ 
 A_t &=&
 \frac{4\Big(\frac{1}{2}-\frac{4}{3}\sin^2\theta_w\Big)}{\sin\theta_w
   \cos\theta_w} \; \Big[I_2(\tau_t,\eta_t)-I_1(\tau_t,\eta_t) \Big] \,, \\
 A_{\xi} &=& \frac{(2\sin^2\theta_w-1)}{\sin\theta_w \cos\theta_w} \;
 I_1(\tau_{\xi},\eta_{\xi}) \,.
\end{eqnarray}
\end{subequations}
The functions $I_1$ and $I_2$ are given by
\begin{subequations}
\begin{eqnarray}
 I_1(\tau,\eta) &=& \frac{\tau \eta}{2(\tau -\eta)} +
 \frac{\tau ^2\eta^2}{2(\tau -\eta)^2}\Big[f(\tau )-f(\eta)\Big]
 +\frac{\tau ^2\eta}{(\tau -\eta)^2}\Big[g(\tau )-g(\eta)\Big]  \,, \\
 I_2(\tau ,\eta) &=& -\frac{\tau \eta}{2(\tau -\eta)}\Big[f(\tau
   )-f(\eta)\Big] \,,  
\end{eqnarray}
\end{subequations}
where the function $f$ has been defined in \Eqn{f}.  Since 
$\tau_x,\eta_x > 1$ for $x=W,t,\xi$, the function $g$ assumes the
following form:
\begin{eqnarray}
 g(a) = \sqrt{a-1}\sin^{-1}\left(\sqrt{1/a}\right) \,.
\end{eqnarray}
In the decoupling limit, the parameter $\kappa$ which appears in
Eqs.\ (\ref{defkappa}), (\ref{h2gg}) and (\ref{h2Zg}) is given by
\begin{eqnarray}
\kappa = \frac{1} {m_\xi^2} (m_A^2 - m_\xi^2 - \frac12 m_h^2) \,.
\label{BGLkappa}
\end{eqnarray}
The appearance of $m_A$ in \Eqn{BGLkappa} is merely an artefact of the
U(1) symmetry in the scalar potential which enforces \Eqn{BGL}.  In
the more general potential of \Eqn{potential}, the expression for
$\kappa$ involves $\lambda_5$, which has nothing to do with $m_A$.
The decoupling behaviour of $\kappa$ for large $m_\xi$ is not then
guaranteed, as also noted in Refs.~\cite{Djouadi:1996yq,
  Arhrib:2004ak}.  However, in the present scenario, unitarity
conditions of \Eqs{uniM1}{uniM2} bound the splitting between heavy
scalar masses, ensuring smooth decoupling of $\kappa$ with increasing
$m_\xi$.  As we have noticed, this splitting is also controlled by the
$T$-parameter.

\begin{figure}[p]
\includegraphics[scale=0.99]{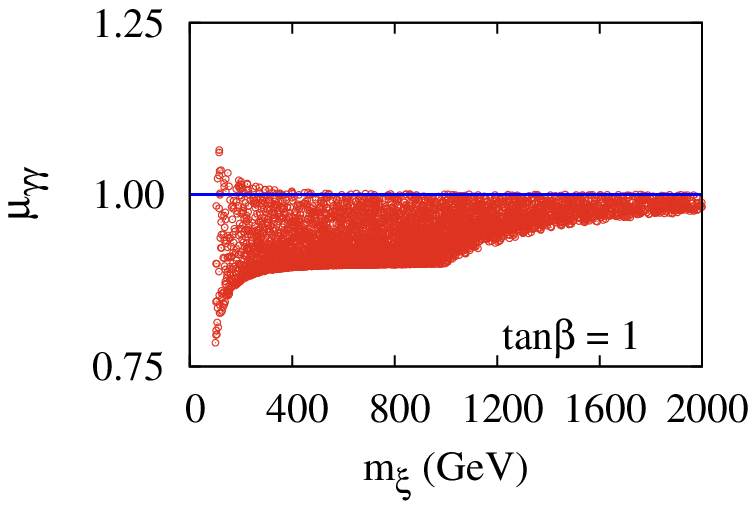} 
\includegraphics[scale=0.99]{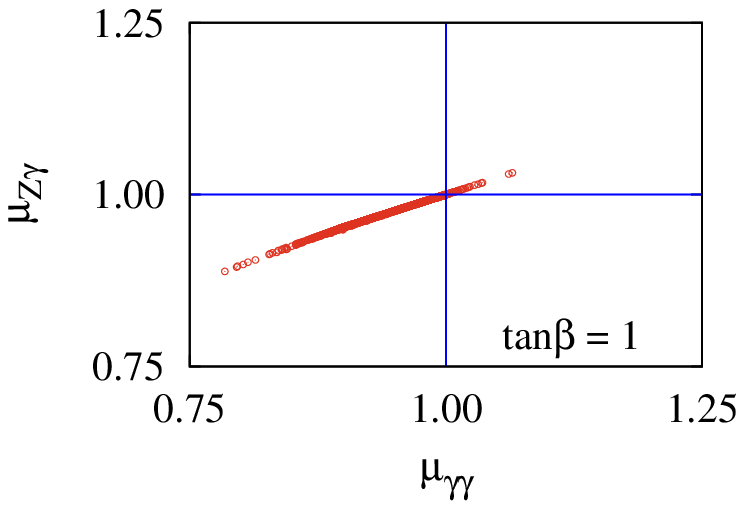} 

\includegraphics[scale=0.99]{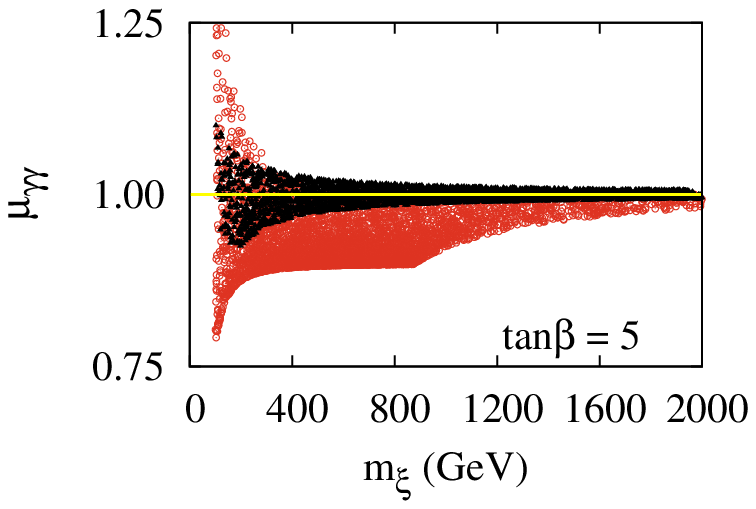}
\includegraphics[scale=0.99]{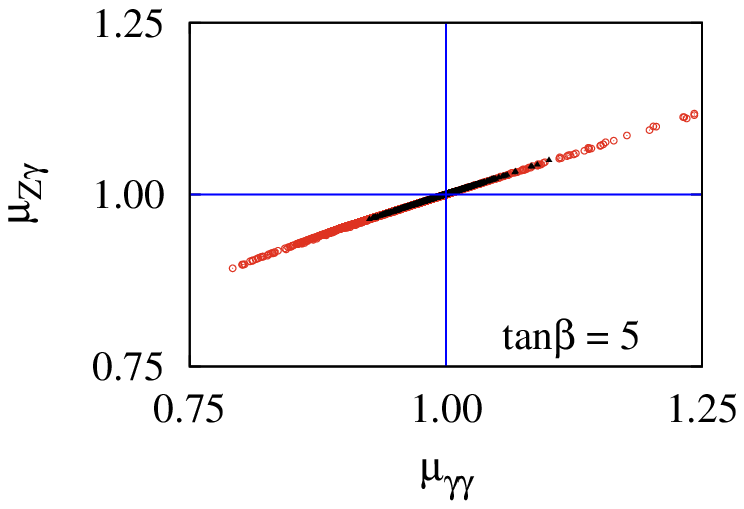}

\includegraphics[scale=0.99]{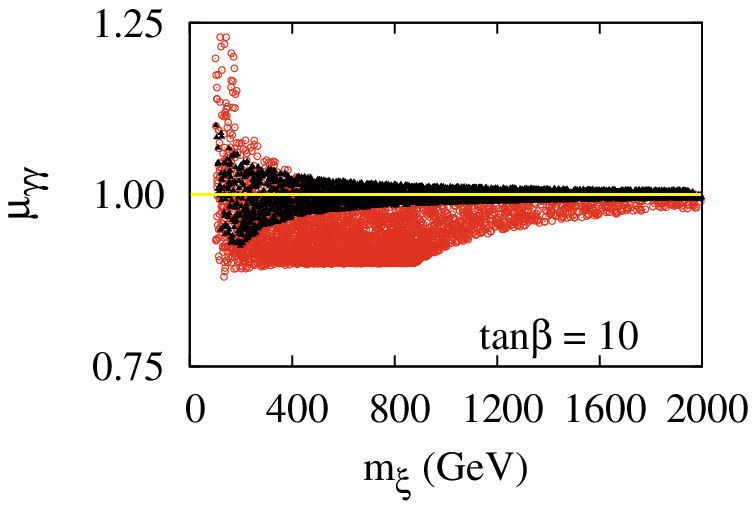}
\includegraphics[scale=0.99]{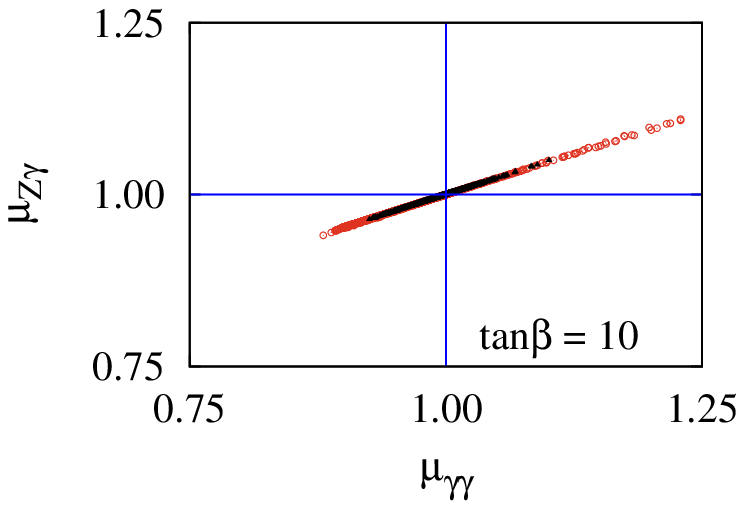}

\caption{The left panels show allowed regions in
  $m_\xi$-$\mu_{\gamma\gamma}$ plane for three values of $\tan\beta$,
  while the right panels show correlation between $\mu_{\gamma\gamma}$
  and $\mu_{Z\gamma}$ for the same choices of $\tan\beta$.  The
  regions shown in red  are allowed by unitarity and stability, while
  the black regions additionally pass the $T$-parameter test.  We put
  $m_\xi > 100$\,GeV and $m_H > m_h$.} 
\label{f:mu}
\end{figure}
In our case, the quantities $\mu_{\gamma\gamma}$ and $\mu_{Z\gamma}$,
defined through the equations
\begin{eqnarray}
 \mu_{\gamma\gamma} &=& {\sigma(pp\to h) \over \sigma^{\rm
     SM}(pp\to h)} \cdot {\mbox{BR} (h \to 
   \gamma\gamma) \over \mbox{BR}^{\rm SM} (h \to \gamma\gamma)} \,,
 \\ 
 \mu_{Z\gamma} &=& {\sigma(pp\to h) \over \sigma^{\rm
     SM}(pp\to h)} \cdot {\mbox{BR} (h \to 
   Z\gamma) \over \mbox{BR}^{\rm SM} (h \to Z\gamma)} \,,
\label{mu}
\end{eqnarray}
assume the following forms:
\begin{eqnarray}
 \mu_{\gamma\gamma} &=&  \frac{\Gamma(h \to \gamma\gamma)}{\Gamma^{\rm
     SM}(h\to \gamma\gamma)} 
 = \frac{\Big|F_W + \frac{4}{3}F_t  +
   \kappa F_{\xi}\Big|^2}{\Big|F_W + \frac{4}{3}F_t \Big|^2} \,,
 \\ 
 \mu_{Z\gamma} &=&  \frac{\Gamma(h \to
  Z\gamma)}{\Gamma^{\rm SM}(h\to Z\gamma)} = \frac{\Big|A_W + A_t  +
  \kappa A_\xi \Big|^2}{ \Big|A_W + A_t \Big|^2} \,.
\end{eqnarray}

In Fig.~\ref{f:mu}, we show the variation of $\mu_{\gamma\gamma}$
against $m_\xi$ and the correlation between $\mu_{\gamma\gamma}$ and
$\mu_{Z\gamma}$ for $\tan\beta=1$, 5, 10.  When we show the variation
of $\mu_{\gamma\gamma}$ with $m_\xi$, we take into consideration all
values of $m_H$ and $m_A$ which are allowed in Fig.~\ref{f:planes}.
As in Fig.~\ref{f:planes}, the red points are those which are allowed
by unitarity and stability constraints, while the superimposed black
points are allowed by the $T$-parameter.  The points allowed by
unitarity prefer suppression in $\mu_{\gamma\gamma}$ compared to the
SM expectation.  Note that large suppressions appear near the lower
end of $m_\xi$ values in Fig.~\ref{f:mu} for $\tan\beta=1$ and 5, but
not for $\tan\beta=10$.  This is because small values of $m_A$,
including $m_A=0$, are allowed for $\tan\beta=1$ and $\tan\beta=5$,
which give sizable negative values of $\kappa$ through \Eqn{BGLkappa}
even for small $m_\xi$.  The correlation between $\mu_{\gamma\gamma}$
and $\mu_{Z\gamma}$ can in principle be used for discriminating new
physics models with increased sensitivity in the future course of the
LHC run.

\section{Conclusions}\label{s:conclu}
In this paper, we have studied quantitative correlation among the
non-standard scalar masses in a class of 2HDM with a global U(1)
symmetry in the potential.  We outline below the salient features of
our analysis.  We derive our constraints from the consideration of
unitarity of scattering amplitudes and the global stability of the
potential.  We additionally sumperimpose the constraints from the
oblique electroweak $T$-parameter on these plots.  We have restricted
our analysis to the decoupling limit which entails a relation between
the parameters $\beta$ and $\alpha$, where the 125\,GeV Higgs boson
has SM-like couplings.  A crucial observation is that when $\tan\beta$
stays close to unity, the CP-odd $A$ can be light, and it is in this
limit that we obtain the maximum deviation (in fact, a suppression) in
the diphoton decay width.  We also observe that for values of
$\tan\beta\sim5$ or larger, all the three non-standard scalar masses
are roughly degenerate.  More specifically, in this limit unitarity
dictates $m_H$ and $m_A$ to be almost equal and $|m_\xi^2-m_H^2|$ to
be small, while the $T$-parameter restricts $|m_\xi-m_H|$ to be very
small.  Another interesting observation is that the charged-Higgs
induced contribution to the diphoton decay amplitude depends on
$(m_\xi^2-m_A^2)/m_\xi^2$, for which the numerator is of order $v^2$,
and therefore the contribution decouples with increasing $m_\xi$.  In
the absence of a global U(1) symmetry in the potential, the parameter
$\lambda_5$ cannot be related to $m_A$, and as a result, the above
decoupling behaviour is not apparent.  It is also important to note
that, thanks to the global U(1) symmetry, unitarity restricts
mass-squared differences and not the individual masses of the
non-standard scalars.

\bigskip

\paragraph*{Acknowledgements\,:} 
MNR thanks the Saha Institute of Nuclear Physics, where this work was
started, for warm hospitality and financial support. The work of MNR
is partially supported by Funda\c{c}\~ao para a Ci\^encia e a Tecnologia
(FCT, Portugal) through the projects \url{CERN / FP / 123580 / 2011},
\url{PTDC / FIS - NUC / 0548 / 2012} and \url{CFTP - FCT} Unit 777
\url{(PEst - OE / FIS / UI0777 / 2013)} which are partially funded
through POCTI (FEDER) and also by the Marie Curie ITN ``UNILHC''
\url{PITN - GA - 2009 - 237920}.  DD thanks the Department of Atomic
Energy, India, for financial support.

\bibliographystyle{unsrt}
\bibliography{2hdm.bib}

\begin{thebibliography}{10}

\bibitem{Aad:2012tfa}
G.~Aad {\it et al.} [ATLAS Collaboration], Phys.\ Lett.\ B {\bf 716} (2012) 1
  [arXiv:1207.7214 [hep-ex]].

\bibitem{Chatrchyan:2012ufa}
S.~Chatrchyan {\it et al.} [CMS Collaboration], Phys.\ Lett.\ B {\bf 716}
  (2012) 30 [arXiv:1207.7235 [hep-ex]].

\bibitem{Chatrchyan:2012jja}
S.~Chatrchyan {\it et al.} [CMS Collaboration], Phys.\ Rev.\ Lett.\ {\bf 110},
  081803 (2013) [arXiv:1212.6639 [hep-ex]].

\bibitem{Ferreira:2011aa}
P.~M.~Ferreira, R.~Santos, M.~Sher and J.~P.~Silva, Phys.\ Rev.\ D {\bf 85},
  077703 (2012) [arXiv:1112.3277 [hep-ph]].

\bibitem{Ferreira:2012my}
P.~M.~Ferreira, R.~Santos, M.~Sher and J.~P.~Silva, Phys.\ Rev.\ D {\bf 85},
  035020 (2012) [arXiv:1201.0019 [hep-ph]].

\bibitem{Swiezewska:2012eh}
B.~Swiezewska and M.~Krawczyk, arXiv:1212.4100 [hep-ph].

\bibitem{Drozd:2012vf}
A.~Drozd, B.~Grzadkowski, J.~F.~Gunion and Y.~Jiang, JHEP {\bf 1305} (2013) 072
  [arXiv:1211.3580 [hep-ph]].

\bibitem{Chang:2012ve}
S.~Chang, S.~K.~Kang, J.~-P.~Lee, K.~Y.~Lee, S.~C.~Park and J.~Song, JHEP {\bf
  1305} (2013) 075 [arXiv:1210.3439 [hep-ph]].

\bibitem{Craig:2013hca}
N.~Craig, J.~Galloway and S.~Thomas, arXiv:1305.2424 [hep-ph].

\bibitem{Chiang:2013ixa}
C.~W.~Chiang and K.~Yagyu, arXiv:1303.0168 [hep-ph].

\bibitem{Jiang:2013pna}
Y.~Jiang, arXiv:1305.2988 [hep-ph].

\bibitem{Chen:2013rba}
C.~-Y.~Chen, S.~Dawson and M.~Sher, arXiv:1305.1624 [hep-ph].

\bibitem{Eberhardt:2013uba}
O.~Eberhardt, U.~Nierste and M.~Wiebusch, JHEP {\bf 07} (2013) 118
  [arXiv:1305.1649 [hep-ph]].

\bibitem{Basso:2013wna}
L.~Basso, A.~Lipniacka, F.~Mahmoudi, S.~Moretti, P.~Osland, G.~M.~Pruna and
  M.~Purmohammadi, PoS Corfu {\bf 2012} (2013) 029 [arXiv:1305.3219 [hep-ph]].

\bibitem{Swiezewska:2013uya}
B.~Swiezewska and M.~Krawczyk, arXiv:1305.7356 [hep-ph].

\bibitem{Branco:2011iw}
G.~C.~Branco, P.~M.~Ferreira, L.~Lavoura, M.~N.~Rebelo, M.~Sher and
  J.~P.~Silva, Phys.\ Rept.\ {\bf 516}, 1 (2012) [arXiv:1106.0034 [hep-ph]].

\bibitem{Glashow:1976nt}
S.~L.~Glashow and S.~Weinberg, Phys.\ Rev.\ D {\bf 15} (1977) 1958.

\bibitem{Paschos:1976ay}
E.~A.~Paschos, Phys.\ Rev.\ D {\bf 15} (1977) 1966.

\bibitem{Grossman:1994jb}
Y.~Grossman, Nucl.\ Phys.\ B {\bf 426} (1994) 355 [hep-ph/9401311].

\bibitem{Joshipura:1990pi}
A.~S.~Joshipura and S.~D.~Rindani, Phys.\ Lett.\ B {\bf 260} (1991) 149.

\bibitem{Antaramian:1992ya}
A.~Antaramian, L.~J.~Hall and A.~Rasin, Phys.\ Rev.\ Lett.\ {\bf 69} (1992)
  1871 [hep-ph/9206205].

\bibitem{Hall:1993ca}
L.~J.~Hall and S.~Weinberg, Phys.\ Rev.\ D {\bf 48} (1993) 979
  [hep-ph/9303241].

\bibitem{Branco:1996bq}
G.~C.~Branco, W.~Grimus and L.~Lavoura, Phys.\ Lett.\ B {\bf 380}, 119 (1996)
  [hep-ph/9601383].

\bibitem{Botella:2009pq}
F.~J.~Botella, G.~C.~Branco and M.~N.~Rebelo, Phys.\ Lett.\ B {\bf 687} (2010)
  194 [arXiv:0911.1753 [hep-ph]].

\bibitem{Botella:2011ne}
F.~J.~Botella, G.~C.~Branco, M.~Nebot and M.~N.~Rebelo, JHEP {\bf 1110} (2011)
  037 [arXiv:1102.0520 [hep-ph]].

\bibitem{D'Ambrosio:2002ex}
G.~D'Ambrosio, G.~F.~Giudice, G.~Isidori and A.~Strumia, Nucl.\ Phys.\ B {\bf
  645} (2002) 155 [arXiv:hep-ph/0207036].

\bibitem{Ferreira:2009jb}
P.~M.~Ferreira and D.~R.~T.~Jones, JHEP {\bf 0908} (2009) 069 [arXiv:0903.2856
  [hep-ph]].

\bibitem{HHGuide}
J. F. Gunion, H. E. Haber, G. Kane, and S. Dawson, The Higgs Hunter's Guide
  (Perseus Publishing, Cambridge, MA, 1990).

\bibitem{Djouadi:1996yq}
A.~Djouadi, V.~Driesen, W.~Hollik and A.~Kraft, Eur.\ Phys.\ J.\ C {\bf 1}
  (1998) 163 [hep-ph/9701342].

\bibitem{Sher:1988mj}
M.~Sher, Phys.\ Rept.\ {\bf 179} (1989) 273.

\bibitem{Gunion:2002zf}
J.~F.~Gunion and H.~E.~Haber, Phys.\ Rev.\ D {\bf 67}, 075019 (2003)
  [hep-ph/0207010].

\bibitem{Lee:1977eg}
B.~W.~Lee, C.~Quigg and H.~B.~Thacker, Phys.\ Rev.\ D {\bf 16} (1977) 1519.

\bibitem{Maalampi:1991fb}
J.~Maalampi, J.~Sirkka and I.~Vilja, Phys.\ Lett.\ B {\bf 265}, 371 (1991).

\bibitem{Kanemura:1993hm}
S.~Kanemura, T.~Kubota and E.~Takasugi, Phys.\ Lett.\ B {\bf 313}, 155 (1993)
  [hep-ph/9303263].

\bibitem{Akeroyd:2000wc}
A.~G.~Akeroyd, A.~Arhrib and E.~-M.~Naimi, Phys.\ Lett.\ B {\bf 490}, 119
  (2000) [hep-ph/0006035].

\bibitem{Horejsi:2005da}
J.~Horejsi and M.~Kladiva, Eur.\ Phys.\ J.\ C {\bf 46}, 81 (2006)
  [hep-ph/0510154].

\bibitem{PDG}
J. Beringer et al. (Particle Data Group), Phys. Rev. D86 (2012) 010001.

\bibitem{He:2001tp}
H.~-J.~He, N.~Polonsky and S.~-f.~Su, Phys.\ Rev.\ D {\bf 64} (2001) 053004
  [hep-ph/0102144].

\bibitem{Grimus:2007if}
W.~Grimus, L.~Lavoura, O.~M.~Ogreid and P.~Osland, J.\ Phys.\ G {\bf 35} (2008)
  075001 [arXiv:0711.4022 [hep-ph]].

\bibitem{Baak:2013ppa}
M.~Baak and R.~Kogler, arXiv:1306.0571 [hep-ph].

\bibitem{Larios:2001ma}
F.~Larios, G.~Tavares-Velasco and C.~P.~Yuan, Phys.\ Rev.\ D {\bf 64} (2001)
  055004 [hep-ph/0103292].

\bibitem{Larios:2002ha}
F.~Larios, G.~Tavares-Velasco and C.~P.~Yuan, Phys.\ Rev.\ D {\bf 66} (2002)
  075006 [hep-ph/0205204].

\bibitem{Cervero:2012cx}
E.~Cervero and J.~-M.~Gerard, Phys.\ Lett.\ B {\bf 712} (2012) 255
  [arXiv:1202.1973 [hep-ph]].

\bibitem{Djouadi:2005gi}
A.~Djouadi, Phys.\ Rept.\ {\bf 457} (2008) 1 [hep-ph/0503172].

\bibitem{Djouadi:2005gj}
A.~Djouadi, Phys.\ Rept.\ {\bf 459} (2008) 1 [hep-ph/0503173].

\bibitem{Arhrib:2004ak}
A.~Arhrib, W.~Hollik, S.~Penaranda and M.~Capdequi Peyranere, Phys.\ Lett.\ B
  {\bf 579} (2004) 361.

\end{thebibliography}

\end{document}